\documentstyle[pra,aps,twocolumn,epsfig]{revtex}
\begin{document}
\flushbottom
\draft
\title{{\bf Field and density coherence of matter-wave fields}}
\author{E.\ V.\ Goldstein, O.\ Zobay, and P.\ Meystre}
\address{Optical Sciences Center, University of Arizona, Tucson,
Arizona 85721\\
{\rm \  }
\\ \medskip}\author{\small\parbox{14.2cm}{\small \hspace*{3mm}
In analogy to Glauber's analysis of optical coherence, we adopt an
operational approach to introduce different classes of atomic coherence
associated with different types of measurements. For the sake of
concreteness we consider specifically
fluorescence, nonresonant imaging and ionization. We introduce definitions 
of coherence appropriate to them, which we call
electronic, density and field coherence, respectively. We illustrate these
concepts in various descriptions of Bose-Einstein condensation, showing that
each of these descriptions makes different implicit assumptions on the
coherence of the system. We also study the impact of elastic collision on
the field and density coherence properties of atom lasers.
\\[3pt]PACS numbers: 03.75.-b, 42.50.Ar, 03.75.Fi, 42.50.Ct}}\address{}
\maketitle
\date{\today}
\maketitle
\narrowtext

\section{Introduction}

Quantum optical coherence theory is based on the factorization properties
of normally-ordered correlation functions of the electric field operator
\cite{Gla65}. This is a direct consequence of the fact that most
optical experiments detect light by absorption,
i.e. by ``removing'' photons from the light field. But the situation is not
so simple in the case of matter-wave fields, and in particular for atomic
de Broglie waves. This is because atomic detectors can work in a number of
different ways: For instance, one can choose to
measure electronic properties of the atoms, or center-of-mass properties,
or both. Measurements may or may not remove atoms from the field, hence the
role of the annihilation operator is not as central as for light fields.

Due to the added complexity of that situation as compared to the
optical case, no unified theory of atomic coherence exists to-date.
The major objective of this paper is to present a step toward
this goal, extending the ideas of optical coherence theory to introduce
several types of matter-waves coherence, in particular ``field coherence''
and ``density coherence.''

In analogy to Glauber's analysis of optical coherence \cite{Gla65}, we adopt
an operational approach where different classes of atomic coherence are
associated with different types of measurements. For the sake of
concreteness we consider specifically three
classes of measurements: fluorescence, nonresonant imaging and ionization.
Section II briefly reviews the outcome of these measurements, and introduces
definitions of coherence appropriate to them. We call them electronic,
density and field coherence, respectively. In the case of bosonic atoms,
field-coherent states are easily seen to correspond to the usual Glauber
coherent states \cite{Gla65}.
In the single-mode case, density-coherent states are
simply given by Fock states, but a more general discussion is required for
the multimode case. Section III illustrates these ideas in various descriptions
of Bose-Einstein condensates. Section IV further develops the concept of
multimode density coherence, which is applied to the case of the atom
laser in Sec.\ V. In this latter example,
we illustrate in particular how elastic collisions, while strongly
modifying the field coherence of the device, have almost no effect on its
density coherence. Finally, Sec.\ VI is a summary and outlook.

\section{Field and density coherence}

A physically appealing operational way to discuss matter-waves
coherence relies on the analysis of specific detection schemes, in complete
analogy with the optical case. As we shall see, this approach naturally
leads to the need to associate different {\em classes of coherence} with
different types of measurements. In addition, it builds useful
bridges between concepts familiar in quantum optics and methods of
traditional manybody theory.

In this paper we consider for concreteness a bosonic atomic Schr\"odinger
field described by a multicomponent field operator
${\hat {\bf \Psi}}({\bf r}, t)$, with components
${\hat \Psi}_i({\bf r}, t), i = 1,2,\ldots$ We normally think of the atoms
comprising this field as adequately described in the Born-Oppenheimer
approximation, so that ${\bf r}$ describes their center-of-mass motion and
the index $i$ labels various electronic states. For bosonic atoms, one
has then
\begin{equation}
[{\hat \Psi}_i({\bf r},t), {\hat \Psi}^\dagger_j({\bf r}',t)] = \delta_{ij}
\delta({\bf r} - {\bf r}') .
\end{equation}

Our goal is to characterize the statistical properties of one or more
components of this field. One familiar method involves the detection of light
fields interacting with the atomic sample, in the hope that properties
of the Schr\"odinger field can be inferred from it. All of laser spectroscopy
relies on this approach, although it is not normally cast in terms of
matter-wave fields. Another approach, which we will find useful in some
respects, involves ionizing atoms from the sample and studying the
properties of the emitted ions or electrons. But this method typically also
relies on the interaction of the atoms with a light field. Hence, a
generic Hamiltonian describing a measurement scheme for the properties
of the atomic Schr\"odinger field is of the form
\begin{equation}
{\cal V} = \sum_{ij} \int d^3 r {\hat \Psi}^\dagger_j({\bf r})
[{\bf d}_{ij}\cdot{\hat {\bf E}}({\bf r},t)] {\hat \Psi}_i({\bf r}) ,
\label{h}
\end{equation}
where ${\hat {\bf E}}({\bf r},t)$ is the electric field operator,
${\bf d}_{ij}$ is the dipole matrix element between electronic states
$i$ and $j$, and we have assumed for simplicity that the electric dipole
approximation gives an adequate description of the atom-field interaction.

We consider the situation where the electromagnetic field consists of a
classically populated field mode of amplitude ${E}_0$, wave vector
${\bf k}_0$ and polarization ${\bbox \epsilon}$, and a series of
weakly excited sidemodes of wave vectors ${\bf k}_\ell$ and polarizations
${\bbox \epsilon}_\ell$. In that case, the Hamiltonian (\ref{h}) becomes
\begin{eqnarray}
{\cal V} &=& \sum_{ij,\ell} \int d^3 r {\hat \Psi}^\dagger_j({\bf r})
\left [\Omega_{ij,0}({\bf r}) e^{i({\bf k}_0\cdot {\bf r} -\omega_0t)}
\right .
\nonumber \\
&+& \left . i\Omega_{ij,\ell} a_\ell e^{i{\bf k}_\ell \cdot{\bf r}}
 \right ]
{\hat \Psi}_i({\bf r}) + H.c. ,
\label{h2}
\end{eqnarray}
where we have introduced the Rabi frequencies
\begin{equation}
\Omega_{ij,0}({\bf r}) =
d_{ij}E_0({\bf r}) ({\bbox \epsilon}_{ij} \cdot {\bbox \epsilon})/\hbar
\end{equation}
corresponding to the classical driving field and the vacuum Rabi frequencies
\begin{equation}
\Omega_{ij,\ell}({\bf r}) =
d_{ij}{\cal E}_\ell({\bf r}) ({\bbox \epsilon}_{ij}
\cdot {\bbox \epsilon}_\ell)/\hbar .
\end{equation}
In these expressions, ${\bbox \epsilon}_{ij}$ is the direction of the
Atomic dipole of magnitude $d_{ij}$ for the $i \leftrightarrow j$
transition, and ${\cal E}_\ell = [\hbar \omega_\ell/2{\bbox \epsilon}_\ell
V]^{1/2}$ is the ``electric field per photon'' in mode $\ell$.

\subsection{Resonance fluorescence}

In resonance fluorescence measurements, one proceeds by shining a laser
quasiresonant with an electronic transition $g \leftrightarrow e$,
and measuring, e.g., the fluorescence spectrum
\begin{equation}
S(\omega) = \int d\tau e^{-i\omega \tau} \langle {\hat E}^-({\bf r}_0, 0)
{\hat E}^+({\bf r}_0, \tau) \rangle  + c.c.
\end{equation}
at the location ${\bf r}_0$ of a photodetector. In that expression,
${\hat E}^+({\bf r}_0, t)$ and ${\hat E}^-({\bf r}_0, t)$ are the
familiar positive and negative frequency parts of the electric field
operator, and we have assumed stationarity to identify the Fourier transform
of the first-order correlation function of the field with its spectrum.
It is well known that except for unimportant retardation effects,
one has that
\begin{equation}
{\hat E}^+({\bf r}_0, t) \propto {\hat \Psi}_e^\dagger ({\bf r}_0, t)
{\hat \Psi}_g ({\bf r}_0, t) ,
\end{equation}
so that
\begin{eqnarray}
S(\omega) &\propto& \int\, d\tau e^{-i\omega \tau} \nonumber \\
&&\langle{\hat \Psi}_g^\dagger ({\bf r}_0, 0)
{\hat \Psi}_e ({\bf r}_0, 0)
{\hat \Psi}_e^\dagger ({\bf r}_0,\tau){\hat \Psi}_g({\bf r}_0,\tau)
\rangle + c.c.
\end{eqnarray}

Resonance fluorescence has been studied in considerable detail, both
experimentally and theoretically. The resonance fluorescence spectrum of
a two-state atom is known to consist of a {\em coherent} and an
{\em incoherent} contribution. For a perfectly monochromatic excitation
laser, the coherent spectrum $S_{coh}$ consists of a delta-function at the
laser frequency, while the incoherent spectrum is the famous Mollow
three-peak spectrum. Mathematically,
\begin{eqnarray}
S_{coh} &\propto&  \int\, d\tau e^{-i\omega \tau} \nonumber \\
&&\langle{\hat \Psi}_g^\dagger ({\bf r}_0, 0)
{\hat \Psi}_e ({\bf r}_0, 0)  \rangle
\langle {\hat \Psi}_e^\dagger ({\bf r}_0,\tau)
{\hat \Psi}_g({\bf r}_0,\tau)
\rangle + c.c.
\end{eqnarray}
and $S_{inc}(\omega) = S(\omega) - S_{coh}(\omega)$. In physical terms, this
means that coherent effects are associated with the {\em factorized}
correlation function
\begin{equation}
\langle{\hat \Psi}_g^\dagger ({\bf r}_0, 0) {\hat \Psi}_e ({\bf r}_0, 0)
\rangle \langle {\hat \Psi}_e^\dagger ({\bf r}_0,\tau)
{\hat \Psi}_g({\bf r}_0,\tau) \rangle .
\end{equation}
These considerations justify defining the ``electronic coherence'' of the
matter field ${\hat {\bf \Psi}}({\bf r}, t)$ in terms of the factorization
properties of normally ordered correlation functions of the
{\em field polarization} operator
\begin{equation}
{\hat \Sigma}_-({\bf r}, t) \equiv
{\hat \Psi}_g^\dagger ({\bf r}, t) {\hat \Psi}_e ({\bf r}, t).
\end{equation}

\subsection{Off-resonant imaging}

In contrast to resonance fluorescence, off-resonant imaging involves
a strongly detuned electromagnetic field interacting with the atoms in the
sample in such a way that it induces only virtual transitions.
After adiabatic elimination of the upper electronic state of the atomic
transition under consideration, the interaction between the Schr\"odinger
field and the radiation field is described to lowest order in the
side-modes by the effective Hamiltonian
\begin{eqnarray}
V & = & \hbar \int d^3r \left [
\frac{|\Omega_0({\bf r})|^2}{\delta_0}
+\sum_\ell
\left (\frac
{\Omega_0({\bf r}) \Omega_\ell^\star}{\delta_0}a_\ell^\dagger
e^{i({\bf k}_0-{\bf k}_\ell)\cdot {\bf r}}\right .\right.
\nonumber\\
&&+\left.\left.
\frac{\Omega_0^\star({\bf r}) \Omega_\ell}{\delta_0} a_\ell
e^{-i({\bf k}_0-{\bf k}_\ell)\cdot {\bf r}}
\right)\right ]{\hat \Psi}^\dagger ({\bf r}) {\hat \Psi} ({\bf r}),
\end{eqnarray}
where the atom-field detuning $\delta_0\equiv \omega_a-\omega_0$ is such
that $|\delta_0|\gg |\Omega_0({\bf r})|$, and we have omitted the index
labeling the ground state component of the Schr\"odinger field, which can
now be considered as scalar.

There are a number of ways in which off-resonant imaging can be
applied to the determination of specific properties of the Schr\"odinger
field. For instance, one can detect interferences between the classical
incident field and scattered light, as in the MIT experiments
\cite{DavMewAnd95}. This results in a signal proportional to the density
$\langle {\hat \rho}({\bf r},t) \rangle $,
where we have introduced the {\em field density} operator
\begin{equation}
{\hat\rho}({\bf r},t) \equiv {\hat \Psi}^\dagger({\bf r},t)
{\hat \Psi}({\bf r},t) ,
\end{equation}
whose expectation value is the local density of the sample.

Alternatively, one can measure the spectrum of the scattered light
in a fashion familiar from resonance fluorescence experiments \cite{Jav95}.
For side-modes initially in a vacuum state, the most important nontrivial
contribution to the fluorescence signal ${\cal F}$ is proportional to the
intensity $|\Omega_0|^2$ of the incident field,
\begin{eqnarray}
& &{\cal F} = \frac {|\Omega_0|^2}{\delta_0^2}
\sum_\ell|\Omega_\ell|^2 \int d^3r d^3 r'
\int_t^{t+\Delta t} d\tau  d\tau'
\nonumber \\
& & \times e^{i(({\bf k}_0 -{\bf k}_\ell)\cdot ({\bf r}-{\bf r}')
-(\omega_0 -\omega_\ell)(\tau-\tau'))}
\langle {\hat \rho}({\bf r}, \tau){\hat \rho}({\bf r}', \tau') \rangle ,
\end{eqnarray}
and hence is sensitive to the second-order correlation function of the
Schr\"odinger field density. Indeed, it can be shown that any measurement
involving the electromagnetic field scattered by the atomic sample under
conditions of off-resonant imaging is determined by correlation functions
of ${\hat\rho}({\bf r},t)$,
\begin{equation}\label{defgd}
D^{(n)}(x_1,\ldots,x_n) =
\langle {\hat \rho}(x_1) \ldots {\hat \rho}(x_n) \rangle ,
\label{dcorr}
\end{equation}
where $x_i \equiv ({\bf r}_i, t_i)$.

In analogy with the optical case, we therefore define a Schr\"odinger field
as being {\em density-coherent to order} ${\cal N}$ if its density correlation
functions $D^{(n)}(x_1, \ldots,x_n)$ factorize for all $n \le {\cal N}$.
>From this definition, it is obvious that single-mode density coherent
states are the familiar number states. But the situation is more complex
for multimode fields, to which we return in Sec.\ IV.

\subsection{Ionization}

\subsubsection{Physical model}

The reason resonance fluorescence and off-resonant imaging yield signals
proportional to correlation functions of {\em }bilinear products of
components of the Schr\"odinger field is of course that the electric dipole
interaction is itself bilinear in the Schr\"odinger field operators.
This raises the question as to whether it is possible to measure
correlation functions of ${\hat {\bf \Psi}}({\bf r}, t)$ itself, as in
the case of optical fields. This can be achieved if
instead of making measurements on the radiation field, one detects the
atoms directly \cite{ThoWan94,ThoWan95}. One possible scheme that achieves
this goal is the ionization method that we now discuss.

Consider a detector consisting of a tightly focussed laser
beam that can ionize atoms by inducing transitions from their ground
electronic level $|g \rangle $ to a continuum level $| i \rangle $.
We are interested in measuring properties of the ground state component
${\hat \Psi}_g({\bf r})$ of this field, which is electric dipole-coupled
to continuum states ${\hat \Psi}_i({\bf r})$. 

In contrast to the preceding
measurement schemes, we are no longer interested in the dynamics of the
light field, whose role is merely to ionize the atoms.
Rather, one extracts information about the state of the Schr\"odinger
field ${\hat \Psi}_g({\bf r}, t)$ by standard atomic physics methods, such
as, e.g., the detection of the quasi-free electrons of the continuum states.
It is therefore sufficient to describe the light field via its 
(possibly time-dependent) classical Rabi
frequencies $\Omega_i$ between the levels $|g \rangle $ and $|i \rangle $,
so that the atom-field interaction reduces to
\begin{equation}
{\cal V} = \hbar \sum_i \int d^3 r \Omega_i({\bf r},t)
{\hat \Psi}_i^\dagger ({\bf r}){\hat \Psi}_g({\bf r})
e^{-i\omega_L t}+ H.c.
\label{vion}
\end{equation}
For ground-state atoms cooled well
below the recoil temperature and tightly focused laser beams, the spatial
size of the atomic wave function is much larger than the laser spot and we
can approximate the electric field ${\bf E}({\bf r})$ by ${\bf E}({\bf r})
\simeq {\bf E} \delta({\bf r}-{\bf r}_0)$, so that Eq. (\ref{vion}) becomes
\begin{equation}
{\cal V} = \hbar \sum_i \Omega_i({\bf r}_0,t)
{\hat \Psi}_i^\dagger ({\bf r_0}){\hat \Psi}_g({\bf r_0})
e^{-i\omega_L t} + H.c.
\label{V}
\end{equation}

We assume for simplicity that the center-of-mass wave functions of the
continuum states of the atoms are well described by plane waves of
momentum ${\bf q}$, so that the single-atom Hamiltonian ${\cal H}_0$ may be
expressed as
\begin{equation}
{\cal H}_0 = {\cal H}_g + \sum_{i{\bf q}} {\cal H}_{i{\bf q}} ,
\end{equation}
where
\begin{equation}
{\cal H}_{i{\bf q}} = \hbar \omega_{iq} b_{i,{\bf q}}^\dagger
b_{i,{\bf q}}.
\end{equation}
Here we expanded ${\hat \Psi}_i({\bf r})$ in plane waves as
\begin{equation}
{\hat \Psi}_i({\bf r}) = \sum_{\bf q} \phi_{{\bf q}}({\bf r})
b_{i,{\bf q}}
\end{equation}
with $[b_{i,{\bf q}}, b^\dagger_{i',{\bf q}'}] = \delta_{{\bf q}{\bf q}'}
\delta_{ii'}$, and $\omega_{iq} = \hbar {\bf q}^2/2M +\omega_i$.

In the following we take the atomic system to be initially in the state
\begin{equation}
|\psi  \rangle = |\{\psi_{i,{\bf q}}\}, \psi_g \rangle .
\end{equation}
To first order in perturbation theory and for CW beams,
the transition probability away from
that state during the time interval $\Delta t$ is
\begin{eqnarray}
& &w  \simeq  \sum_{i,{\bf q},i',{\bf q}'} |\Omega_i({\bf r}_0)|^2
\int_t^{t+\Delta t} d\tau \int_t ^{t+\Delta t} d\tau'
\nonumber\\
& &\times \left [
e^{i\omega_L(t-t')}\langle \psi_g |{\hat \Psi}_g^\dagger({\bf r}_0, \tau)
{\hat \Psi}_g ({\bf r}_0,\tau') |\psi_g \rangle\right.
\nonumber\\
& &\times
\langle \{\psi_{i,{\bf q}} \}|{\hat \Psi}_i({\bf r}_0, \tau)
|\{\phi_{i',{\bf q}'}\}\rangle
\langle \{\phi_{i',{\bf q}'}\}|
{\hat \Psi}_i^\dagger({\bf r}_0,\tau') |\{\psi_{i,{\bf q}}\} \rangle
\nonumber\\
& &+ e^{-i\omega_L(t-t')}\langle \psi_g |{\hat \Psi}_g({\bf r}_0, \tau)
{\hat \Psi}_g^\dagger ({\bf r}_0,\tau') |\psi_g \rangle
\nonumber\\
& &
\times \langle \{\psi_{i,{\bf q}} \}|{\hat \Psi}_i^\dagger({\bf r}_0, \tau)
|\{\phi_{i',{\bf q}'}\}\rangle
\langle \{\phi_{i',{\bf q}'}\}|\left.
{\hat \Psi}_i({\bf r}_0,\tau') |\{\psi_{i,{\bf q}}\} \rangle \right ],
\label{w}
\end{eqnarray}
where the sum is over all final states $|\{\phi_{i',{\bf q}'}\}\rangle$
in the excited state manifold. In this expression, we have neglected
contributions involving the product of two creation or
annihilation operators, a result
of the implicit assumption that any atom in the continuum will be removed
from the sample instantaneously. In addition, we explicitly carried out
the sum over all final states of the ground-state field, but not for the
excited fields manifold. This is because we want to allow for the possibility
of selective detection of the ionized atoms. Following Ref. \cite{Gla65},
this can be easily achieved by replacing the sum over final states in
Eq.\ (\ref{w}) by a weighted sum
\begin{equation}
\sum_{i',{\bf q}'} \rightarrow \sum_{i',{\bf q}'} {\cal R}(i', {\bf q}') ,
\label{sum}
\end{equation}
where ${\cal R}(i', {\bf q}')$ is the detector sensitivity to atoms in
state $|\phi_{i', {\bf q}'} \rangle.$ In practice, we have in mind
energy-selective detectors, ${\cal R}(i', {\bf q}') \rightarrow
{\cal R}(\omega)$, and the degeneracy of the levels must then of course
be accounted for.

There is a fundamental distinction between the situation at hand and
Glauber's photodetection theory, because in the present case both
the detected and detector fields consist of
matter waves. There is a complete symmetry between these two fields so far,
and their roles are interchangeable.
In order to break this symmetry and to truly construct a detector,
we now make a series of assumptions on the state of the detector fields
${\hat \Psi}_i({\bf r}, t)$. Physically, this amounts to making a statement
about the way the detector is prepared prior to a measurement. Specifically,
we assume that all atoms are in the ground state, $\Psi_i({\bf r}_0,0)
|\{\psi_{i, {\bf q}}\} \rangle =0 $, and that any atom in an
ionized state will be removed from the sample instantaneously, as already
mentioned. In that case, the second term in Eq.\ (\ref{w}) vanishes.

\subsubsection{Energy-selective detectors}

To illustrate this result, we consider the situation of
energy-selective detectors, and discuss the limits of
{\em narrowband} and {\em broadband} detection \cite{Gla65,CohDupGry92}.
In the first case the detector bandwidth $\Delta E_d$ is assumed to be much
narrower than the energy width $\Delta E_g$ of the ground state
Schr\"odinger field, which is determined solely by the spread in
center-of-mass momentum (temperature) since all atoms occupy the same
internal state. The reverse is true in the second case.

For a narrowband detector, the substitution of Eq.\ (\ref{sum}) into
Eq.\ (\ref{w}) yields readily
\begin{equation}
r_{nb}(\omega)\propto
 \int_0^{\Delta t} d\tau e^{-i(\omega-\omega_L) \tau}
G_{A}(t,t+\tau;{\bf r}_0,{\bf r}_0).
\label{int}
\end{equation}
Here $\hbar \omega$ is the energy of the registered photoelectrons, and
we introduced the ionization rate $r_{nb}(\omega)=w_{nb}(\omega)/ \Delta t$
and the normally ordered first-order correlation function of the ground
state Schr\"odinger field
$G_{A}(t,t';{\bf r}_0,{\bf r}_0)=
\langle {\hat \Psi}_g^\dagger({\bf r}_0, t)
{\hat \Psi}_g ({\bf r}_0,t')  \rangle.$
The only time intervals that significantly contribute to the integral
(\ref{int}) are such that $\Delta t\simeq 1/\Delta \omega_g$. For
large enough detection times this integral can therefore safely be extended
to infinity. In that limit, the detector measures the Fourier component of
the atomic correlation function $G_{A}(t,t';{\bf r}_0,{\bf r}_0)$. For
stationary fields, the Wiener-Khintchine theorem implies that tuning
the detector sensitivity ${\cal R}(E)$ yields the spectrum of
the Schr\"odinger field ${\hat \Psi}_g({\bf r},0)$.

In the case of broadband detection, in contrast, the energy distribution
$\Delta E_d$ of the ionized states is much broader than
$\Delta E_g$. This situation can be realized e.g. by
exciting the ground state with a broadband laser pulse and detecting the
resulting electrons (or ions) with a broadband detector ${\cal R}(E)
\simeq$ constant. Assuming that the spectrum of the ground atoms
Schr\"odinger field is centered at $\bar \omega$ we find
\begin{equation}
r_{bb} \simeq \eta({\bf r}_0) G_{A}(t,t;{\bf r}_0,{\bf r}_0),
\end{equation}
where we have introduced in prevision of the following discussion the
``detector cross-efficiency''
\begin{eqnarray}
& &\eta({\bf r}_1,{\bf r}_2)=\sum_i \Omega_i({\bf r}_1)
\Omega_i^\star({\bf r}_2)
\nonumber \\
& &\int_0 ^{\Delta t} d\tau
\langle\hat{\Psi}_i({\bf r}_2,t+\tau)\hat{\Psi}^\dagger_i({\bf r}_1,t)\rangle
e^{-i(\bar{\omega}-\omega_L) \tau} ,
\label{eta}
\end{eqnarray}
from which the usual detector efficiency is simply recovered as
$\eta({\bf r}_0) \equiv \eta({\bf r}_0,{\bf r}_0)$.
As expected, a broadband detector is not able to resolve any spectral
feature of the Schr\"odinger field, and only measures the local atomic
density, like off-resonant imaging.\\

\subsubsection{Higher-order correlations}

The detection of higher-order correlations of the Schr\"odinger
field can be achieved by a straightforward generalization of the ionization
detector. For instance, second-order coherence measurements can be carried
out by focussing the laser at two locations ${\bf r}_1$ and ${\bf r}_2$, in
which case
$$
{\cal V} = \hbar\sum_{\mu=1,2} \sum_j \Omega_j({\bf r}_\mu)
{\hat \Psi}^\dagger_j({\bf r}_\mu) {\hat \Psi}_g({\bf r}_\mu)
e^{-i\omega_L t} + H.c.
$$
The joint probability to ionize an atom at ${\bf r}_1$ and another one at
${\bf r}_2$ is then
\begin{eqnarray}
& &w_2 \simeq \sum_{\{j_i\}\{{\bf q}_i\}}
\int_{t}^{t+\Delta t}d\tau_1\int_{t}^{t+\Delta t}d\tau_2
\int_{t}^{t+\Delta t}d\tau_3\int_{t}^{t+\Delta t}d\tau_4
\nonumber\\
& &\times
e^{-i\omega_L(\tau_1+\tau_2-\tau_3-\tau_4)}
\Omega_{j_1}^\star({\bf r}_1)\Omega_{j_2}^\star({\bf r}_2)
\Omega_{j_3}({\bf r}_2)\Omega_{j_4}({\bf r}_1)
\nonumber\\
&&\times\langle {\hat \Psi}_{j_1}({\bf r}_1, \tau_1)
{\hat \Psi}_{j_2}({\bf r}_2, \tau_2)
{\hat \Psi}_{j_3}^\dagger({\bf r}_2, \tau_3)
{\hat \Psi}_{j_4}^\dagger({\bf r}_1, \tau_4)\rangle
\nonumber\\
& &\times
\langle {\hat \Psi}_g^\dagger({\bf r}_1, \tau_1){\hat \Psi}_g^\dagger
({\bf r}_2, \tau_2){\hat \Psi}_g({\bf r}_2, \tau_3)
{\hat \Psi}_g({\bf r}_1, \tau_4) \rangle.
\label{final}
\end{eqnarray}
It involves two detected atoms, hence it is now
necessary to properly account for the quantum statistics of the measured
particles. For this purpose, we describe the ionized atoms as ion-electron
pairs, whereby the electrons are described by the creation and annihilation
operators $c_{{\bf k}s}^\dagger$ and $c_{{\bf k}s}$ satisfying
Fermi commutation relations $[c_{{\bf k}s},c^\dagger_{{\bf k}'s'}]_+=
\delta_{ss'}\delta_{{\bf k}{\bf k}'}$. Here $s$ labels
the electron spin and ${\bf k}$ its momentum. We similarly introduce
ion creation and annihilation operators
$a_{{\bf k} s}^\dagger, a_{{\bf k} s}$, also
satisfying Fermi commutation relations (for bosonic atoms.)
For a spin-zero atom, the atomic mode operators $b_{j,{\bf q}}$
can be expressed in terms of the ion and electron operators as
\begin{eqnarray}
& &b_{j,{\bf q}}\equiv|j{\bf q}\rangle\langle 0|
=\sum_{{\bf k}{\bf k}'s s'}|{\bf k}{\bf k}'s s'\rangle
\langle {\bf k}{\bf k}'s s'|j{\bf q}\rangle\langle 0|
\nonumber\\
& &=\sum_{{\bf k} s}\varphi_j({\bf k}) a_{{\bf q}s}c_{{\bf k}-s}
=\sum_s a_{{\bf q}s}c_{j -s},
\label{atom}
\end{eqnarray}
where $\varphi_j({\bf k})$ are electron wave functions in ${\bf k}$-space,
$c_{j s}=\sum_{\bf k} \varphi_j({\bf k})c_{{\bf k}s}$
and we have assumed that the center-of-mass wave function is
$e^{i{\bf q}\cdot{\bf r}}$ with ${\bf r}$ being the ion
(or atomic center-of-mass) position. Due to spin conservation
the values of electron and ion spins are clearly opposite.

Substituting this result into Eq.(\ref{final}) yields, in the case of
broadband detection,
\begin{eqnarray}
& &w_2=\eta({\bf r}_1)\eta({\bf r}_2)
\int_t^{t+\Delta t}d\tau_1 \int_t^{t+\Delta t} d\tau_2
\nonumber\\
& &\times
\langle \Phi_g^\dagger({\bf r}_1,\tau_1)\Phi_g^\dagger({\bf r}_2,\tau_2)
\Phi_g ({\bf r}_2,\tau_2)\Phi_g({\bf r}_1,\tau_1)\rangle
\nonumber\\
& &+\eta({\bf r}_1,{\bf r}_2)\eta({\bf r}_2,{\bf r}_1)
\int_t^{t+\Delta t} d\tau_1 \int _t^{t+\Delta t}d\tau_2
\nonumber\\
& &\times\langle\Phi_g^\dagger({\bf r}_1,\tau_1)
\Phi_g^\dagger({\bf r}_2,\tau_2)\Phi_g
({\bf r}_2,\tau_1)\Phi_g({\bf r}_1,\tau_2)\rangle]
\nonumber\\
& & +\eta_x({\bf r}_1,{\bf r}_2)\int _t^{t+\Delta t}d\tau_1
\nonumber\\
& &\times\langle\Phi_g^\dagger({\bf r}_1,\tau_1)
\Phi_g^\dagger({\bf r}_2,\tau_1)\Phi_g
({\bf r}_2,\tau_1)\Phi_g({\bf r}_1,\tau_1)\rangle\},
\label{final1}
\end{eqnarray}
where the detector sensitivity $\eta_x({\bf r}_1,{\bf r}_2)$
to processes involving electron exchange is
\begin{eqnarray}
& &\eta_x({\bf r}_1,{\bf r}_2)=\int_t^{t+\Delta t}
d \tau_2 \int_t^{t+\Delta t} d \tau_3 \int_t^{t+\Delta t} d\tau_4
\nonumber\\
& &\times e^{-i\omega_L(\tau_1+\tau_2-\tau_3-\tau_4)}
\sum_{\alpha \beta {\mbox{\boldmath {$\kappa$}}}{\bf q}}
\nonumber\\
& &\times\left [e^{i(\omega_\kappa(\tau_1-\tau_3)
+\omega_q(\tau_2-\tau_4)+\omega_\alpha(\tau_1-\tau_4)
+\omega_\beta(\tau_2 -\tau_3))}\right.
\nonumber\\
& &\times |\Omega_\alpha({\bf r}_1)|^2 |\Omega_\beta({\bf r}_2)|^2
\phi_{\mbox{\boldmath {$\kappa$}}}
^\star({\bf r}_1)\phi_{\mbox{\boldmath {$\kappa$}}}
({\bf r}_2)
\phi_{\bf q}^\star({\bf r}_2)\phi_{\bf q}({\bf r}_1)
\nonumber\\
& &+\Omega_\alpha^\star({\bf r}_1)\Omega_\alpha({\bf r}_2)
\Omega_\beta^\star({\bf r}_2) \Omega_\beta({\bf r}_1)
|\phi_{\mbox{\boldmath {$\kappa$}}}({\bf r}_1)|^2|\phi_{\bf q}({\bf r}_2)|^2
\nonumber\\
& & \times
\left. e^{i(\omega_\kappa(\tau_1-\tau_4)+\omega_q(\tau_2-\tau_3)
+\omega_\alpha(\tau_1-\tau_3)+ \omega_\beta(\tau_2 -\tau_4))}
\right ].
\end{eqnarray}

The first term in Eq. (\ref{final1}) is familiar from double photodetection,
with the usual exchange contributions from the {\em detected} field.
The second term is an additional exchange term
due to the fact that the detector field is a single Schr\"odinger
field. Its origin is the interference of the {\em detector} field at points
${\bf r}_1$ and ${\bf r}_2$. It is absent in conventional
photodetection theory, a result of the implicit assumption that the
two detectors used are distinguishable. \footnote{
A similar comment can be made
about the position measurement scheme discussed in Refs.
\cite{ThoWan94,ThoWan95}. In that case, the absence of the detector exchange
contribution can be traced back to the assumption that the set of states
excited at each location are distinguishable. While this
approximation is usually reasonable, it becomes
questionable in situations involving quantum degenerate gases such as
Bose-Einstein condensates.}
Finally, the term proportional to
$\eta_x$ results from the fact that electrons do not know from which
atom they originate. We note that these last two terms can be eliminated
by using a gated detection scheme \cite{CohDupGry92}
that eliminates the contribution of the
exchange terms in the detector field. In practice, such gating
can be achieved by using non-overlapping short laser pulses to ionize
the atoms. In that case, the ionization scheme simply yields normally
ordered correlation functions of the Schr\"odinger field
\begin{equation}
G^{(n)}(x_1,\ldots,x_n) =
\langle  {\hat \Psi}^\dagger(x_1) \ldots {\hat \Psi}^\dagger(x_n)
 {\hat \Psi}(x_n) \ldots {\hat \Psi}(x_1) \rangle ,
\end{equation}
in complete analogy with the optical situation. This also justifies
defining a Schr\"odinger field as {\em field coherent to order } ${\cal N}$
if its normally ordered correlation functions $G^{(n)}$  factorize for
all $n \le {\cal N}$.

\section{Example: Bose-Einstein condensation}

\subsection{Hartree description}

To illustrate the ideas developed in the preceding section, we consider
a quantum-degenerate $N$-particle bosonic system described by the state
\begin{equation}
|\psi(t)\rangle_N=\frac{1}{\sqrt{N!}}
\int d\{{\bf r}_i\} f_N(\{{\bf r}_i\},t)\Pi_i 
\hat{\Psi}^\dagger({\bf r}_i)|0\rangle,
\end{equation}
where the $N$-body wave function $f_N(\{{\bf r}_i\},t)$ 
is totally symmetric in
its arguments. If the sample forms a condensate, it is described to an
excellent degree of approximation by a Hartree wave function, whereby
the $N$-body wave function $f_N(\{{\bf r}_i\},t)$
factorizes as a product of the form $f_N(\{{\bf r}_i\},t)=\Pi_i
\phi_N({\bf r}_i,t)$.
That is, all atoms in the condensate are described by the same Hartree
wave function $\phi_N({\bf r},t)$ and the $N$-particle state reduces to
\begin{equation}
|\psi(t)\rangle_N=\frac{1}{\sqrt{N!}}
\left(\int d{\bf r} \phi_N({\bf r},t)
\hat{\Psi}^\dagger({\bf r})\right)^N |0\rangle.
\end{equation}
The equation of motion for $\phi_N({\bf r},t)$ is obtained from the
variational principle
\begin{equation}
\frac{\delta}{\delta \phi^\star_N({\bf r})}\left[_N\langle\psi|i\hbar
\frac{\partial}{\partial t}-{\cal H}|\psi\rangle_N\right]=0,
\label{varpri}
\end{equation}
where ${\cal H}$ is a manybody Hamiltonian.

We expand then the Schr\"odinger field as 
\footnote{Note that this
representation of the Schr\"odinger field operator is different
from the Bogoliubov description, a consequence of the fact that the system
is {\em not} taken to be in a field coherent state.}
\cite{WriWalGar96}
\begin{equation}
\hat{\Psi}^\dagger({\bf r})=\phi^\star_N({\bf r})
a^\dagger +\delta {\hat \psi}^\dagger({\bf r}),
\end{equation}
where $a^\dagger$ is the creation operator for a particle with the Hartree
wave function $\phi_N(x)$ and the operator $\delta {\hat \psi}^\dagger
({\bf r})$ creates particles in all other states.
Further assuming that  the condensate state is stationary
\begin{equation}
\phi_N({\bf r},t)=\phi_N({\bf r})e^{-i\mu_N t/\hbar}
\end{equation}
with $\mu_N$ being the chemical potential, the state of the system is
simply
\begin{equation}
|\psi(t)\rangle_N=\frac{e^{-i\mu_N t/\hbar}}
{\sqrt{N!}}(a^\dagger)^N|0\rangle
\equiv |N\rangle ,
\label{numb}
\end{equation}
where we have
used the orthogonality of the condensate state to all other
modes, i.e. $\int d{\bf r}\phi_N({\bf r})\delta\psi^\dagger({\bf r})|0\rangle
=0$.
Hence the Hartree approximation is equivalent to the assumption that
the condensate is in a number state of the self-consistent Hartree ``mode.''
In this description the condensate is in a density coherent state
(\ref{numb}). Indeed, $|\psi(t)\rangle_N$ is easily seen to be
an eigenstate of the field density operator $\hat{\rho}({\bf r})$,
\begin{equation}
\hat{\rho}({\bf r})|\psi(t)\rangle_N=N\rho({\bf r})|\psi(t)\rangle_N
\end{equation}
where $\rho({\bf r})
=|\phi_N({\bf r})|^2$ is the local Hartree density of the condensate.
It is straightforward to see that this state is density-coherent to all
orders. However, just like single-mode number states of the electromagnetic
field, it does not exhibit {\em field} coherence past first order coherence.

\subsection{Wave packet description}

For small systems containing a finite number of atoms the assumption
that all particles are in the condensate state is actually not realistic.
Rather, the particle number in a condensate fluctuates,
so that only the mean number of atoms in the condensate is known.
In this picture it is appropriate to describe the condensate as a wave
packet in Fock space, and assume that the particle
number distribution is sharply peaked around ${\bar N}$. Following Ref.
\cite{WriWalGar96} we assume a Poissonian particle
number distribution and thus, the state of a condensate reads
\begin{equation}
|\psi(t)\rangle=e^{-\bar{N}}\sum_N \frac{\bar{N}^{N/2}}{N!}
e^{-iN\mu_N t/\hbar}(a^\dagger)^N|0\rangle .
\label{wpack}
\end{equation}
In contrast to a pure condensate state with fixed number of atoms
in the Hartree ground state, this state can be characterized
with a nonzero ``order parameter'' defined by
\cite{WriWalGar96}
\begin{equation}
\Phi({\bf r},t)\equiv\langle \psi(t)|\Psi({\bf r})|\psi(t)\rangle=
\sqrt{\bar{N}}\phi_{\bar {N}}({\bf r})e^{-i\mu t/\hbar}{\cal F}_N(t),
\label{phixt}
\end{equation}
where $\mu=\mu_{\bar N}+{\bar N}\mu_{\bar N}'$, $\mu_{\bar N}'\equiv
\partial {\mu_N}/\partial N |_{N={\bar N}}$ and
\begin{equation}
{\cal F}_N(t)=e^{-\bar{N}}\sum_N\frac{\bar{N}^{(N-1)}}{(N-1)!}
e^{-2i\mu_{\bar{N}}'(N-{\bar N}) t/\hbar}.
\end{equation}
This order parameter exhibits periodical collapses and
revivals due to the dispersion of the chemical potential over the particle
number variance. Physically, this is a consequence of the fact that while
the initial state of the condensate is field (Glauber) coherent, it does not
remain so in the course of time. Hence, the state (\ref{wpack})
is neither field, nor density coherent.\\

Interestingly, it turns out that the {\em one-time}
field coherence functions measured
in ionization experiments (Sec.\ II.C) are factorizable.
Indeed, the first order field coherence function in this case reads
$$G^{(1)}({\bf r};0)=\bar N |\phi_{\bar N}({\bf r})|^2$$ and the second
order field coherence function is
$$G^{(2)}({\bf r},{\bf r}';0)={\bar N}^2 |\phi_{\bar N}({\bf r})|^2
|\phi_{\bar N}({\bf r}')|^2,$$
so that the normalized second order coherence function $g^{(2)}(0)
=1$.

\subsection{Spontaneous symmetry breaking description}

As a final possible description of the condensate, we consider the standard
spontaneous symmetry breaking approach whereby the Schr\"odinger field
operator is replaced by a $c$-number \cite{LifPit80},
\begin{equation}
\hat{\Psi}({\bf r},t)\rightarrow\Phi({\bf r},t).
\label{cnumb}
\end{equation}
This description is equivalent to the assumption that the state of
the system is an eigenstate of the Schr\"odinger field operator
$\hat{\Psi}({\bf r},t)|\psi\rangle=\Phi({\bf r},t)|\psi\rangle$,
or, in other words,
that the condensate is in a Glauber coherent state.
This state can be obtained from the previous wave packet description in
the thermodynamic limit when the dispersion of $\mu$ over the particle
number variance becomes negligible and $\mu_N$ can be approximated
by $\mu_{\bar{N}}$. In that case
\begin{equation}
\Phi({\bf r},t)|\psi\rangle
=\sqrt{\bar{N}}e^{-i\mu_{\bar{N}}t} \phi_{\bar{N}}({\bf r})|\psi\rangle
\end{equation}
the condensate remains in a Glauber coherent state at all times.
It is
field coherent, but not density-coherent since for that state
one finds easily that
\begin{equation}
D^{(2)}({\bf r},{\bf r}';0) = D^{(1)}({\bf r})D^{(1)}({\bf r}')
+ D^{(1)}({\bf r}) \delta({\bf r}-{\bf r}') .
\label{dglau}
\end{equation}

These results illustrate how different descriptions of the condensate make
different implicit assumptions about its coherence properties. These models
are amenable, at least in principle, to experimental tests. We note finally
that all descriptions reviewed in this section become equivalent in the
thermodynamic limit as far as their coherence properties are concerned,
despite the fact that their order parameters are different. 

\section{Multimode density correlations}
\subsection{Density coherent states}
In this section we further develop the notion of density coherence
and introduce density-coherent states for a multimode Schr\"odinger
field.\footnote{The discussion in this section can also be
applied to a more general class of density correlation functions. To
this end, consider a one-particle observable $B$ with (discrete or continuous)
eigenvalues $b_n$ and eigenvectors $|b_n\rangle$. One might choose
$B={\bf p}$, for example. In analogy to ${\hat
\Psi}^{\dagger}({\bf r},t)$ and ${\hat \rho}({\bf r}, t)$ one can construct
operators ${\hat \Psi}^{\dagger}(b_n,t)$ and ${\hat \rho}(b_n, t)$. In
terms of these operators one defines the density correlation functions
according to Eq.~(\ref{defgd}). The subsequent derivations then apply
similarly to these correlations functions. In case of $B$ having a
discrete spectrum the mathematical problems connected with the
normalizability of the density-coherent states do not arise.}

>From the definition (\ref{defgd}) of
the $n$-th order density correlation function, we have readily that
\begin{equation}
D^{(n)}(x_1,\dots,x_n)=D^{(n)}(x_n,\dots,x_1)^* .
\end{equation}
If all $x_i$ are taken at the same time $D^{(n)}$ is real. Similarly
to the case of field correlation functions \cite{WalMil94}
one can derive the inequality
\begin{eqnarray}
& & D^{(2n)}(x_1,\dots,x_n,x_n,\dots,x_1) \times\nonumber \\
& & D^{(2m)}(y_m,\dots,y_1,y_1,\dots,y_m)
\geq \nonumber \\
& & |D^{(m+n)}(x_1,\dots,x_n,y_1,\dots,y_m)|^2 .
\end{eqnarray}
The state of a system is said to be ${\cal N}$-th order density
coherent if all density correlation functions up to order ${\cal N}$ 
factorize, i.e.,
\begin{equation}\label{defdc}
D^{(n)}(x_1,\dots,x_n)= D^{(1)}(x_1)...D^{(1)}(x_n),\quad n\leq {\cal N},
\end{equation}
where we recall that $D^{(1)}(x)$ is nothing but the expectation value of
the Schr\"odinger field density, see Eq. (\ref{dcorr}).

Disregarding mathematical rigor, states which are density coherent if all
$x_i$ are taken at the same time can be constructed as follows. Define
\begin{equation}
|y_m,\dots,y_1\rangle = \frac 1 {\sqrt{{\cal C}}}
\hat\Psi^{\dagger}(y_m)\dots\hat\Psi^{\dagger}(y_1) |0\rangle
\end{equation}
with ${\cal C}=\langle 0|\hat\Psi(y_1)\dots\hat\Psi(y_m)
\hat\Psi^{\dagger}(y_m)\dots\hat\Psi^{\dagger}(y_1)
|0\rangle$.\footnote{The normalization constant ${\cal C}$
is not well defined mathematically, of course.
This is due to the continuous nature of the eigenvalue
spectrum of the position operator $\hat{\bf r}$. If one considered
density correlation functions for a discrete operator $B$ (as outlined in
the previous footnote) these problems would not arise.}
This state describes a system of $m$ particles which are localized at
points $y_1,\dots,y_m$. For this state the density correlation functions
are of the form (\ref{defdc}) with
\begin{equation}
D^{(1)}(x)=\sum_{i=1}^m \delta(x-y_i)
\end{equation}
which is a direct consequence of the relation
\begin{equation}
\hat\rho(x)|y_m,\dots,y_1\rangle = \left[\sum_{i=1}^m \delta(x-y_i)\right]
|y_m,\dots,y_1\rangle.
\end{equation}
Due to dispersion the states $|y_m,\dots,y_1\rangle$ will not remain
density coherent in the course of time.\footnote{However, the momentum
eigenstates will be ``momentum density coherent'' at all times for free
particles, i.e. if $H={\bf p}^2/2m$.}

For the sake of comparison we also compute the equal-time first-order
correlation function for a general two-particle state
\begin{equation}
|\phi\rangle=\frac 1 {\sqrt{2}}\int dx_1 dx_2\, \phi(x_1,x_2)
\hat\Psi^{\dagger}(x_1) \hat\Psi^{\dagger}(x_2)|0\rangle
\end{equation}
with symmetric two-particle wave function $\phi(x_1,x_2)$. Denoting the
marginal distribution $\int dx_2 |\phi(x_1,x_2)|^2$ by $p(x_1)$ one obtains
\begin{equation}
D^{(1)}(x_1,x_2)=2|\phi(x_1,x_2)|^2+2p(x_1)\delta(x_1-x_2),
\end{equation}
compare with Eq. (\ref{dglau}). This correlation function is not
factorizable in general.

\subsection{Thermal fields}
A simple system which can be discussed in the context of density
correlations is provided by the thermal multimode Schr\"odinger
field. Here for the sake of simplicity we consider an ideal thermal Bose
gas of atoms without condensate fraction (i.e. above the Bose condensation
transition temperature $T_c$) \footnote{ The calculation of the second-order
field coherence function for the thermal Bose gas
with both interactions and condensate fraction present can be found
in \cite{DodClaEdw97}.}
in which case the state of the system is described
by the density operator
\begin{equation}
\hat{\rho}_T=e^{-\sum_k (\epsilon_k-\mu)a_k^\dagger a_k/k_BT}/Z,
\end{equation}
where $\epsilon_k$ is the eigenenergy of the $k-$th mode, $k_B$
the Boltzmann constant, $T$ is temperature, $\mu$ the chemical potential
and $Z$ the partition function. The equal-time
first-order field coherence function of this system is
\begin{equation}
G^{(1)}(x)= \mbox{Tr}[\hat{\rho}_T \hat{\Psi}^\dagger(x)\hat{\Psi}(x)]
= \sum_k{\bar {n_k}} |\phi_k(x)|^2,
\end{equation}
where we have expanded the Schr\"odinger field operator in terms of mode
annihilation operators in the usual way as
$\hat{\Psi}(x)=\sum_k \phi_k(x)a_k^\dagger$, and the mean number of
atoms in mode $k$ is  ${\bar n}_k =Tr(\hat{\rho}_T a_k^\dagger a_k)$.
Similarly, the equal-time second-order field coherence function is
\begin{eqnarray}
G^{(2)}(x_1,x_2)&=&\mbox{Tr}[\hat{\rho}_T \hat{\Psi}^\dagger(x_1)
\hat{\Psi}^\dagger(x_2)\hat{\Psi}(x_2)\hat{\Psi}(x_1)]\nonumber\\
& =&\sum_{k_1k_2} {\bar n}_{k_1}{\bar n}_{k_2}
(|\phi_{k_1}(x_1)|^2|\phi_{k_2}(x_2)|^2 \nonumber\\
& &+\phi^\star_{k_1}(x_1)\phi_{k_1}(x_2)\phi^\star_{k_2}(x_2)
\phi_{k_2}(x_1))
\end{eqnarray}
These correlation functions are
not factorizable. The one-point normalized second-order coherence
function is
\begin{equation}
g^{(2)}(x)\equiv G^{(2)}(x,x)/G^{(1)}(x)G^{(1)}(x)=2.
\end{equation}
The equal-time second-order density coherence function is
\begin{eqnarray}
D^{(2)}(x_1,x_2)&=&\mbox{Tr}[\hat{\rho}_T \hat{\Psi}^\dagger(x_1)\hat{\Psi}
(x_1)\hat{\Psi}^\dagger(x_2)\hat{\Psi}(x_2)\nonumber \\
&=&G^{(2)}(x_1,x_2)+G^{(1)}(x_1) \delta(x_1-x_2)
\end{eqnarray}
and is likewise not factorizable. Thus the thermal Schr\"odinger field is
neither field nor density coherent.

\section{Field versus density coherence in a dynamical system:
The binary-collisions atom laser}

In order to illustrate how differently the field and density coherence of a
Schr\"odinger field can behave in a dynamical system, we compute these
properties for a model of a binary collisions atom laser. We show in
particular that elastic collisions, while being extremely detrimental to the
field coherence, and hence the linewidth of the laser, have almost no
influence on its density coherence.

\centerline{\psfig{figure=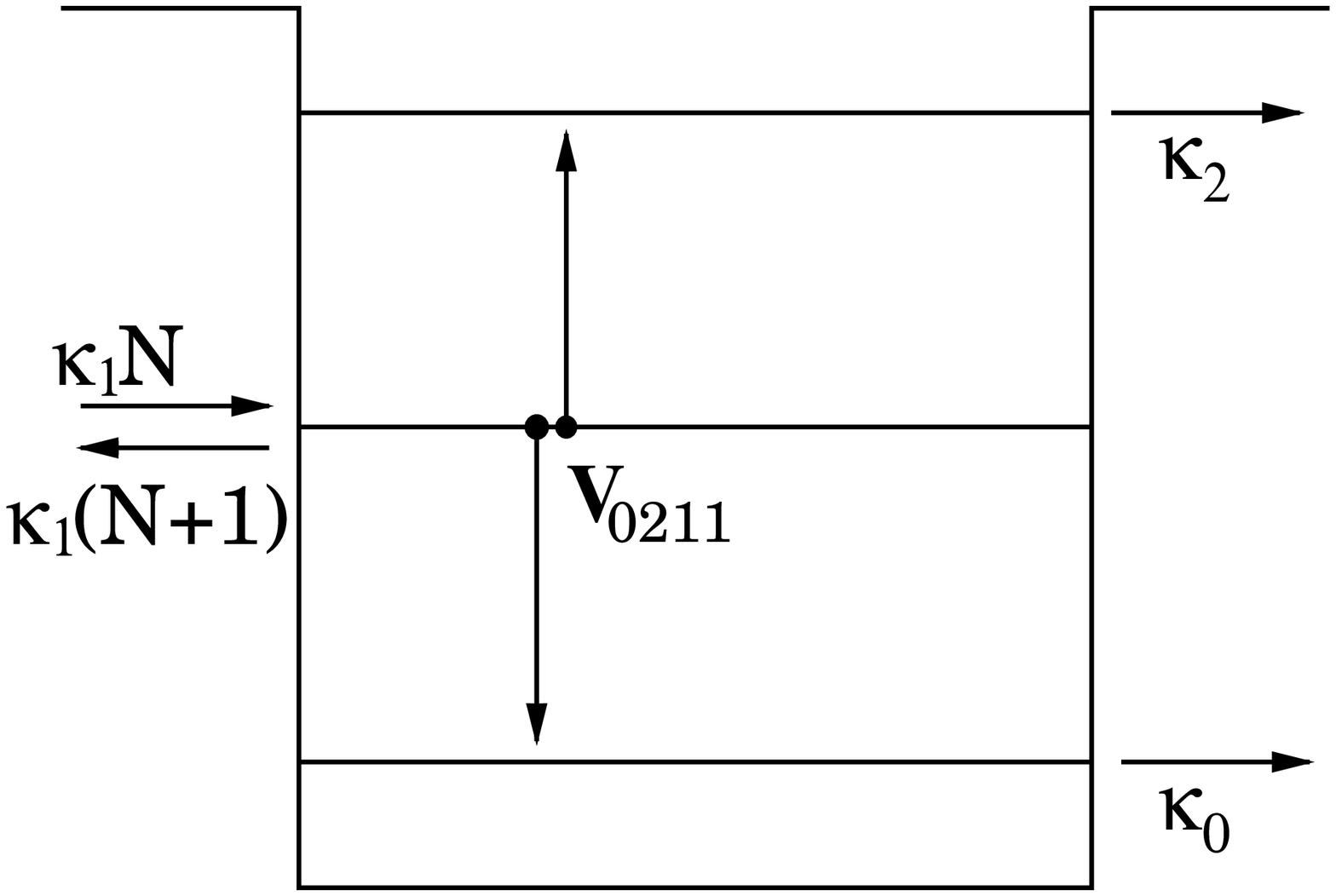,width=8.6cm,clip=}}
\begin{figure}
\caption{Schematic three-mode atom laser model.}
\label{fig11}
\end{figure}

This binary collisions atom laser has been investigated in various publications
\cite{GuzMooMey96,WisMarWal96,HolBurGarZol96,MooMey971,MooMey972,ZobMey97}.
One considers a resonator for atoms, realized e.g.\ by optical fields,
in which only three atomic center-of-mass modes are taken into account
explicitly (cf.\ Fig.~\ref{fig11}).
Bosonic atoms in their ground electronic state are
pumped into an atomic resonator level of ``intermediary'' energy (mode 1).
They then undergo binary
collisions which take one of the atoms involved to the tightly bound
laser mode 0, whereas the other one is transferred to the heavily damped
loss mode 2. This latter atom leaves the resonator quickly, thereby
providing the irreversibility of the pumping process. A macroscopic
population of the laser mode can build up as soon as the influx of
atoms due to pumping compensates for the losses induced by the damping.

For the master equation of this atom laser model one makes the ansatz
\begin{eqnarray} \label{al3m}
\dot{W}& = & -i[H_0+H_{col},W]+\kappa_0{\cal D}[a_0]W+\kappa_1(N+1){\cal D}
[a_1]W \nonumber \\
 & & +\kappa_1 N {\cal D}[a_1^{\dagger}]W + \kappa_2{\cal D}[a_2]W
\end{eqnarray}
with $\hbar=1$.
In this equation, the second quantized formalism is used in which each
center-of-mass atomic mode is associated with an annihilation operator
$a_i$, and $W$ denotes the atomic density operator\footnote{Since we
consider ground state atoms only, they
are fully described by their center-of-mass quantum numbers.}.
The free Hamiltonian is given by
$$H_0=\sum_{i=0,1,2} \omega_i a_i^{\dagger} a_i,$$
$\omega_i$ being the mode frequencies. For the following we set
$\omega_0+\omega_2=2\omega_1$. This condition can be fulfilled, e.g., if
the atomic cavity is realized by time-modulated optical fields
\cite{MooMey971}.

The operation of the atom laser relies on binary collisions between
the atoms in the resonator. The general form of the corresponding
interaction Hamiltonian is
\begin{equation}\label{hcol}
H_{col}=\sum_{i\leq j,k\leq l} V_{ijkl} a^{\dagger}_i a^{\dagger}_j
a_k a_l
\end{equation}
where $V_{ijkl}$ are the matrix elements of the two-body interaction
responsible for the collisions. However, for the present
investigation we restrict our attention to the simplified form
\begin{equation}\label{hcolres}
H_{col} = V_{0211}a^{\dagger}_0 a^{\dagger}_2 a_1 a_1 + V_{1102}
a^{\dagger}_1 a^{\dagger}_1 a_0 a_2 + V_{0000} a^{\dagger}_0
a^{\dagger}_0 a_0 a_0
\end{equation}
in which (besides the pumping collisions) only collisions involving
ground state atoms are retained. The damping rates of the
cavity modes are given by the coefficients $\kappa_i$, and the strength of
the external pumping of mode 1 is characterized by the parameter $N$,
which is the mean number of atoms to which mode 1 would equilibrate in
the absence of collisions. The superoperator ${\cal D}$ is of the Lindblad
form and is defined by
\begin{equation}
{\cal D}[c] P = cPc^{\dagger}-\textstyle{\frac 1 2}(c^{\dagger} c P +
P c^{\dagger} c)
\end{equation}
with arbitrary operators $c$ and $P$.

In order to achieve a sufficiently high degree of irreversibility it is
necessary that $\kappa_2$ is much larger than the damping rates of the
other modes. This suggests to adiabatically eliminate
this mode, an approximation that leads to the simplified master equation
\cite{GuzMooMey96,WisMarWal96,HolBurGarZol96}
\begin{eqnarray} \label{al2m}
\dot{\rho} & = & -i[H_{c},\rho]+\kappa_0{\cal D}[a_0]\rho
+\kappa_1(N+1){\cal D} [a_1]\rho \nonumber \\
 & & +\kappa_1 N {\cal D}[a_1^{\dagger}]\rho + \Gamma {\cal D}
[a_0^{\dagger}a_1^2]\rho.
\end{eqnarray}
Equation (\ref{al2m}) is written in the interaction picture with respect
to $H_0=\omega_0 a^{\dagger}_0 a_0 + \omega_1 a^{\dagger}_1 a_1$, and the
reduced density matrix $\rho$ is
$\rho=\mbox{Tr}_{\mbox{mode 2}}[W]$. The reduced collision Hamiltonian
$H_c$ is
\begin{equation}
H_c = V_{0000}a_0^\dagger a_0^\dagger a_0 a_0,
\end{equation}
and $\Gamma=4|V_{0211}|^2/\kappa$. Consistently
with Ref.~\cite{WisMarWal96} we call the limiting cases
$\Gamma \ll \kappa_0$ and $\Gamma \gg \kappa_0$ the weak and strong
pumping regimes, respectively.

\subsection{Linearized fluctuation analysis}

In order to obtain analytical approximations for the correlation
functions
\begin{equation}
G^{(2)}_j(\tau)=\langle a_j^{\dagger}(0) a_j^{\dagger}(\tau)
a_j(\tau) a_j(0)\rangle
\end{equation}
and 
\begin{equation}
D^{(2)}_j(\tau)=\langle a_j^{\dagger}(\tau) a_j(\tau) a_j^{\dagger}(0)
a_j(0)\rangle
\end{equation}
in the two-mode sytem a linearized
fluctuation analysis can be performed \cite{WalMil94,MeySar91}.
To this end the master equation (\ref{al2m}) is converted to a
Fokker-Planck equation using the $P$-function representation as described
in \cite{WisMarWal96}. This equation can be transformed to polar
coordinates $\alpha_j=\sqrt{n_j}e^{i\phi_j}$, where $\alpha_j$ denotes the
complex amplitudes originally appearing in the Fokker-Planck equation
\cite{Gar83,Gar91}. This leads to stochastic differential equations
\begin{eqnarray}
dn_0&=&[\Gamma n_1^2 (n_0+1) - \kappa_0 n_0] dt + dS_{n_0}, \label{sden0} \\
dn_1&=&[\kappa_1(N-n_1)-2\Gamma n_1^2(n_0+1)]dt + dS_{n_1}, \label{sden1} \\
d\phi_0&=&[-V_{0000}(2n_0-1) -V_{0101}n_1]dt + dS_{\phi_0}, \label{sdephi0} \\
d\phi_1&=&[-V_{1111}(2n_1-1) -V_{0101}n_0]dt + dS_{\phi_1}. \label{sdephi1}
\end{eqnarray}
The correlation matrix ${\bf D}$ for the stochastic forces 
$d{\bf S}^T=(dS_{n_0}, dS_{n_1},dS_{\phi_0},dS_{\phi_1})$ is given by
\begin{equation} \label{diffmat}
\hspace*{-2mm}
\left(\begin{array}{cccc} 2\Gamma n_1^2 n_0 & -2\Gamma n_1^2 n_0
& -2V_{0000}n_0 & -V_{0101} n_0 \\
-2\Gamma n_1^2 n_0 & 2\kappa_1 Nn_1-2\Gamma n_1^2 n_0 & -V_{0101}n_1 &
-2V_{1111}n_1 \\
-2V_{0000}n_0 & -V_{0101}n_1 & \Gamma n_1^2/(2n_0) & \Gamma n_1/2 \\
-V_{0101} n_0 & -2V_{1111}n_1 & \Gamma n_1/2 & \frac{\kappa_1 N}
{2n_1} + \frac{\Gamma n_0}{2} \end{array} \right).
\end{equation}
In the limit $n_0\gg 1$ one obtains from Eqs.~(\ref{sden0})
and (\ref{sden1}) the above-threshold semiclassical steady-state
populations \cite{WisMarWal96}
\begin{eqnarray}
\bar{n}_0&=&\frac 1 2 \frac{\kappa_1}{\kappa_0}(N-\bar{n}_1),\label{ssn0} \\
\bar{n}_1&=&\sqrt{\frac{\kappa_0}{\Gamma}}, \label{ssn1}
\end{eqnarray}
the threshold condition being $N > \sqrt{\kappa_0/\Gamma}$. The drift
terms in Eqs.~(\ref{sden0}) -- (\ref{sdephi1}) and the correlation
matrix ${\bf D}$ do not depend on the phases $\phi_j$. This means that
the time evolution of the atom numbers $n_j$ is {\em not} influenced by the
phase dynamics and is thus completely determined by Eqs.~(\ref{sden0})
and (\ref{sden1}) alone. To proceed further we introduce the fluctuation
variables $\delta n_j=n_j-\bar{n}_j$. In the linear approximation their
time evolution is given by
\begin{equation}
d\,\delta {\bf n}= - {\bf k}\, \delta {\bf n} dt + d{\bf S}_n
\end{equation}
where $\delta {\bf n}^T=(\delta n_0,\delta n_1)$ and
the matrix ${\bf k}$ is obtained by linearizing the drift terms in
Eqs.~(\ref{sden0}) and (\ref{sden1}) around the steady-state values
$\bar{n}_j$. The correlation matrix ${\bf D}_n$ for the stochastic forces
$d{\bf S}_n^T=(dS_{n_0},dS_{n_1})$ is given by the upper left
$2\times2$-minor of the matrix ${\bf D}$ of Eq.~(\ref{diffmat}) after
replacing $n_j$ by $\bar{n}_j$.

\subsection{Second-order correlation function $G^{(2)}$}

In the linear approximation the steady-state second-order correlation 
function for mode $j$ is given by \cite{WalMil94,MeySar91}
\begin{equation}\label{g2lin}
G^{(2)}_{j}(\tau)=(e^{-{\bf k}\tau}{\bbox \sigma})_{jj}+\bar{n}_j^2
\end{equation}
($\tau\ge 0$) where ${\bbox \sigma}$ is the steady-state covariance matrix
\begin{equation}
{\bbox \sigma}=\frac{\Delta {\bf D}_n + ({\bf k}-\Sigma){\bf D}_n
({\bf k}-\Sigma)^T}{2\Sigma\Delta}
\end{equation}
\centerline{\psfig{figure=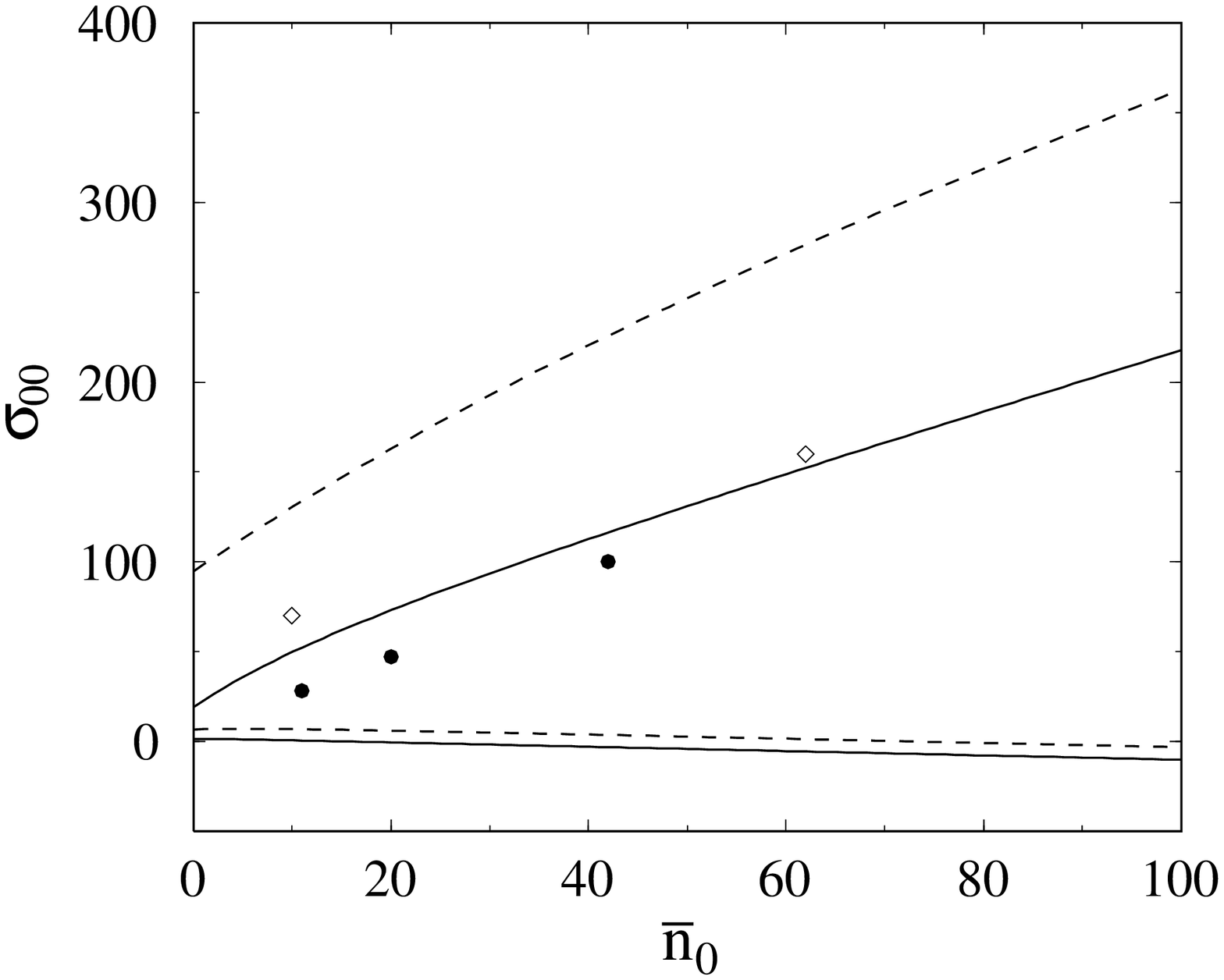,width=8.6cm,clip=}}
\begin{figure}
\caption{Dependence of $\protect{\bbox \sigma}_{00}$ on $\bar{n}_0$. The curves
are calculated using Eq.~(\protect\ref{varn0}) for parameter values
$\Gamma=0.07\kappa_0$ (ascending curves) and $\Gamma=15\kappa_0$
(descending curves), $\kappa_1=20\kappa_0$ (full),
$\kappa_1=100\kappa_0$ (dashed). The single points show numerical
results for $\Gamma=0.07\kappa_0$, $\kappa_1=20\kappa_0$
($\bullet$) and $\kappa_1=100\kappa_0$ ($\diamond$), respectively.}
\label{fig21}
\end{figure}

\noindent with $\Delta=\det {\bf k}$ and $\Sigma=\mbox{Tr\,}{\bf k}$ (we
use the numbers 0,1 as indices for the $2\times 2$-matrices).
For the matrix elements of ${\bbox \sigma}$ one obtains the expressions
\begin{eqnarray}
{\bbox \sigma}_{00}&=&[\bar{n}_0^2(2\kappa_0-\kappa_0/\bar{n}_1)
+\bar{n}_0
(\kappa_1 \bar{n}_1+\kappa_0+\kappa_1) \nonumber \\
& & +\kappa_1^2 \bar{n}_1/(4\kappa_0)]/
(4\Gamma \bar{n}_1 \bar{n}_0+\kappa_1), \label{varn0} \\
{\bbox \sigma}_{11}&=&\frac{\kappa_0(2\bar{n}_1-1)\bar{n}_0 + \kappa_1
\bar{n}_1^2+\kappa_0 \bar{n}_1}{4\kappa_0\bar{n}_0/\bar{n}_1+\kappa_1},
\label{varn1}\\
{\bbox \sigma}_{01}&=&{\bbox \sigma}_{10}=-\bar{n}_1/2.
\end{eqnarray}
In the far-above-threshold limit $\bar{n}_0 \gg 1$ the covariances can
be approximated as ${\bbox \sigma}_{00}\approx\frac 1 2 (\bar{n}_1-
\frac 1 2) \bar{n}_0$ and ${\bbox \sigma}_{11}\approx\frac 1 2
(\bar{n}_1-\frac 1 2) \bar{n}_1$, respectively. For $\bar{n}_1> 1/2$,
i.~e.\ in the weak-pumping regime, the second-order correlation
functions thus show bunching, whereas for $\bar{n}_1< 1/2$, i.~e.\
in the strong-pumping regime, they are anti-bunched. The normalized
second-order correlation $g^{(2)}_0(0)=G^{(2)}_0(0)/\bar{n}_0^2$ goes to
1 with $1/\bar{n}_0$ for large $\bar{n}_0$. In contrast, $g^{(2)}_1(0)
\to 3/2-1/(4\bar{n}_1)$. For the case of $\bar{n}_1 \gg 1$ these
results are in agreement with the conclusions of
Ref.~\cite{WisMarWal96}. To illustrate the physical contents of
Eqs.~(\ref{varn0}) and (\ref{varn1}) in Fig.~\ref{fig21} the dependence of
${\bbox \sigma}_{00}$ on $\bar{n}_0$ is shown for the weak and
strong-pumping regime and different values of $\kappa_1$.

An expansion of $e^{-{\bf k}\tau}$ in the parameter $\bar{n}_0/
\bar{n}_1$ yields the approximate result
\begin{equation}\label{g2tapp}
G^{(2)}_{j}(\tau)={\bbox \sigma}_{jj}e^{-q_j\tau}+\bar{n}_j^2
\end{equation}
for the time dependence of the correlation functions (\ref{g2lin}).
Thereby, the $q_j$ are the two eigenvalues of the matrix ${\bf
k}$. For $\bar{n}_0/\bar{n}_1>1$ they are approximately given by
\begin{eqnarray}\label{defq}
q_0&=&\frac{4\kappa_0^2}{4\kappa_0+\kappa_1 \bar{n}_1/\bar{n}_0},\\
q_1&=&\kappa_1 + 4\kappa_0\bar{n}_0/\bar{n}_1
\end{eqnarray}
Their inverses can be interpreted as the relevant timescales for the
dynamics of the atom number fluctuations. Numerical comparison to
Eq.~(\ref{g2lin}) shows that the
approximation (\ref{g2tapp}) is very accurate, in general. The time
dependence of the second-order correlation functions is thus purely
exponential. The correlation function $G^{(2)}_{0}$ decays on a timescale
of the order of $\kappa_0$ whereas the time evolution of $G^{(2)}_{1}$
is much more rapid.

\subsection{Density-correlation function $D^{(2)}$}

Using Eq.~(10.5.28) of
Ref.~\cite{Gar83} we find that in the linear approximation the steady-state
density-correlation function $D^{(2)}_j$ for mode $j$ is given by
\begin{equation}
D^{(2)}_j(\tau)=\bar{n}_j(e^{-{\bf k}\tau})_{jj}+G^{(2)}_{j}(\tau).
\end{equation}
From Eqs.~(\ref{varn0}) and (\ref{varn1}) it follows that
$D^{(2)}_j(0) > \bar{n}_j^2$.
In analogy to Eq.~(\ref{g2tapp}) one obtains the expression
\begin{equation}\label{d1t}
D^{(2)}_j(\tau)=(\bar{n}_j + {\bbox \sigma}_{jj})e^{-q_j\tau}+
\bar{n}_j^2
\end{equation}
for the explicit time dependence of $D^{(2)}_j$. The intensity
fluctuation spectrum, i.e.\ the Fourier transform of $D^{(1)}_j(\tau)-
\bar{n}_j^2$, is thus a Lorentzian.

It should be noted that all results of this section are independent of
the rate of elastic collisions (as quantified by $V_{0000}$) between
atoms in the laser mode. This means that in the two-mode description
the atom laser is
second-order coherent (at least in the sense of $g^{(2)}_0(\tau)\approx
1$ for all $\tau$) even if $V_{0000}$ is large. In such cases
the first-order correlation function $G^{(1)}$ will decay very rapidly so
that the laser is not first-order coherent, and is characterized by a large
linewidth.

\subsection{Numerical examples}

As discussed in detail in Ref.~\cite{MooMey972} the master
equations (\ref{al3m}) and (\ref{al2m}) may be solved numerically with the
help of quantum Monte Carlo techniques \cite{DumParZolGar92,MolCasDal93}.
The correlation functions $G^{(2)}$ and $D^{(2)}$ can be calculated
within this approach according to the description given in
Ref.~\cite{MolCasDal93}. In the following we compare some numerical
results with the analytical predictions of Secs.~IV.B and C.

{\it Strong-pumping regime.} Figure \ref{fig31}(a) depicts a typical result
for the calculation of $D^{(2)}_0$ and $G^{(2)}_0$ in the
\centerline{\psfig{figure=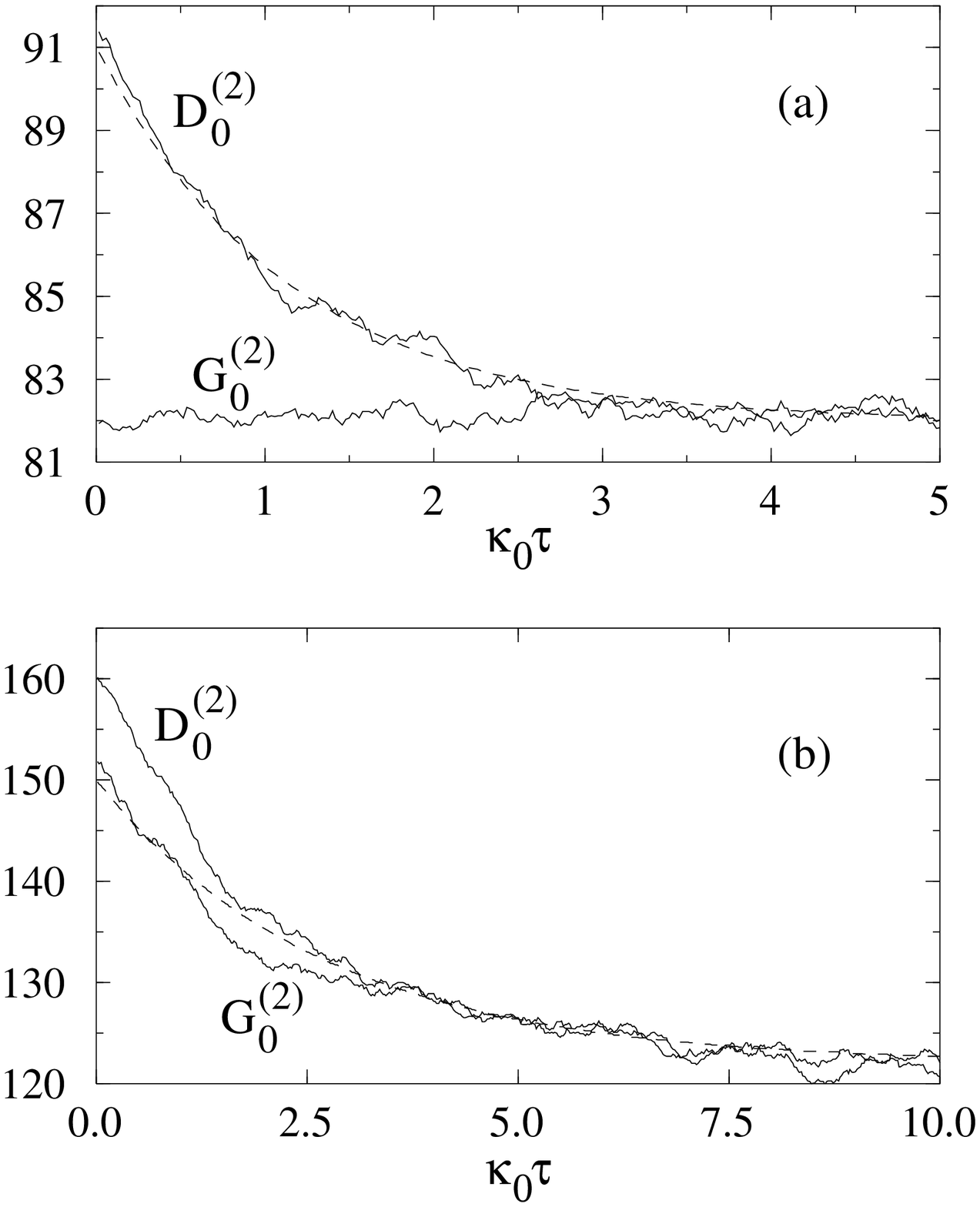,width=8.6cm,clip=}}
\begin{figure}
\caption{Numerical calculation of $D^{(2)}_0(\tau)$ and $G^{(2)}_0
(\tau)$ for parameter values $\Gamma=15\kappa_0$, $\kappa_1=20\kappa_0$,
$N=1.2$ (a) and $\Gamma=0.07\kappa_0$, $\kappa_1=20\kappa_0$, $N=3.9$ (b).
Dashed curves: exponential decay with corresponding time constants $1/q_0$.}
\label{fig31}
\end{figure}

\noindent strong-pumping regime. In this example, the parameters
$\Gamma=15\kappa_0$,
$\kappa_1=20\kappa_0$, and $N=1.2$ were chosen yielding
a numerical equilibrium population of $\bar{n}_0=9.0$
(9.4 analytically).
As should be expected, $D^{(2)}_0(0)-G^{(2)}_0(0)\approx\bar{n}_0$.
The behavior of $D^{(2)}_0(\tau)$ is very well approximated
by an exponential decay with time constant $1/q_0$, as predicted in
Eq.~(\ref{d1t}). From the behavior of $G^{(2)}_0(\tau)$ it can be
inferred that $|g^{(2)}_0(\tau)-1| \ll 1$ for all $\tau$.
However, even after a very large
number of Monte Carlo simulations numerical noise prevents any
further details of the time dependence of $G^{(2)}_0(\tau)$ to be
identified. This observation applies to most calculations in the
strong-pumping regime. In particular, it could not be unambiguously determined
whether antibunching actually occurs for larger values of $\bar{n}_0$. On the
other hand, these numerical results are compatible with the fact that
Eq.~(\ref{varn0}) predicts a small value of $|g^{(2)}_0(\tau)-1|$ (compare
with Fig.~\ref{fig21}).

It should be noted that a small amount of antibunching in $G^{(2)}_0$
could be observed in calculations for the three-mode model in cases in
which $V_{0000}$ is large compared to $V_{0211}$. Under these
conditions, however, the main effect to be observed is a significant
decrease in the equilibrium population $\bar{n}_0$.

{\it Weak-pumping regime.} In contrast to the previous
\centerline{\psfig{figure=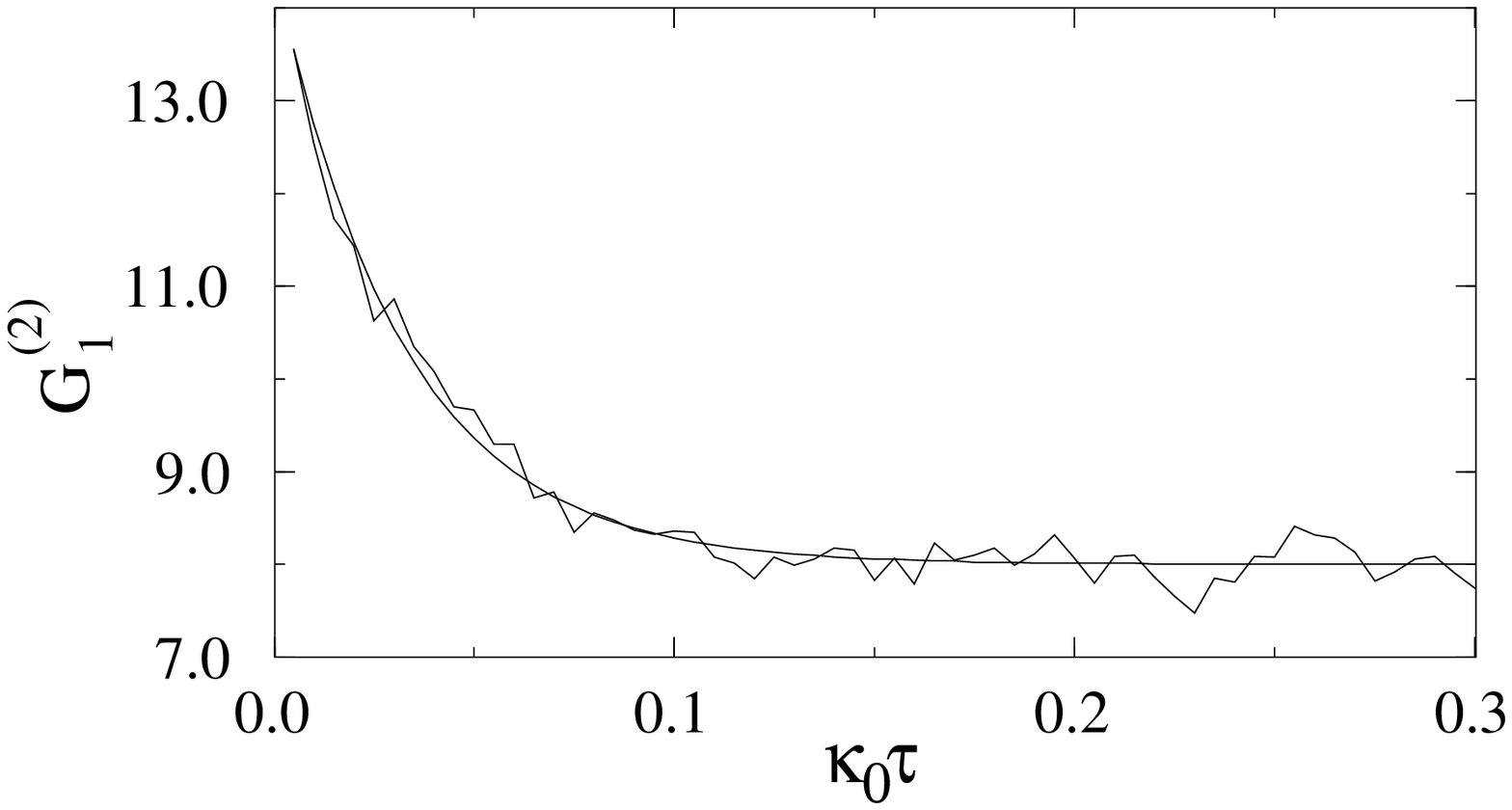,width=8.6cm,clip=}}
\begin{figure}
\caption{Numerical calculation of $G^{(2)}_1 (\tau)$ for
parameter values $\Gamma=0.07\kappa_0$, $\kappa_1=20\kappa_0$, $N=3.9$.
Dashed curve: exponential decay with time constants $1/q_1$.}
\label{fig32}
\end{figure}

\noindent case, for the
weak-pumping regime it can be expected on the grounds of the linear
analysis that $g^{(2)}_0(0)$ is significantly different from 1 if
$\bar{n}_0$ is not too large (cf.\ Fig.~\ref{fig21}). As exemplified in
Fig.~\ref{fig31}(b) this prediction is indeed confirmed by the numerical
calculations. There the correlation functions $D^{(2)}_0$ and
$G^{(2)}_0$ are shown for the parameter values $\Gamma=0.07\kappa_0$,
$\kappa_1=20\kappa_0$, and $N=3.9$. Unfortunately, the quantitative
results of the linear analysis in the weak-pumping
regime are not very accurate for low values of $\bar{n}_0$ (to which the
numerical computations have to be restricted due to time constraints).
For example, for the above parameters Eq.~(\ref{ssn0}) yields
$\bar{n}_0=1.2$ which is much smaller than the numerical value of 11.
However, Eq.~(\ref{defq}) still constitutes a good approximation to the
decay rate of the correlation functions if it is used with the
numerically determined values of $\bar{n}_0$ and $\bar{n}_1$. This is
demonstrated by the dashed curve in Fig.~\ref{fig31}(b) which depicts an
exponential decay with a time constant calculated in this way. A similar
agreement was also found in other examples.

In Fig.~\ref{fig21} the
results of several numerical calculations of ${\bbox \sigma}_{00}$ are
shown as a function of the numerical value of $\bar{n}_0$ for the parameters
$\Gamma=0.07\kappa_0$, $\kappa_1=20\kappa_0$
($\bullet$) and $\kappa_1=100\kappa_0$ ($\diamond$).
The numerical values of ${\bbox \sigma}_{00}$ should be
understood as having an error margin of at least $\pm 10$ -- 15\%. The
results depicted agree in order of magnitude with the predictions of
Eq.~(\ref{varn0}) and also demonstrate a dependence of ${\bbox \sigma}_{00}$
on $\bar{n}_0$ and $\kappa_1$ similar to the analytical one.

{\it Second-order correlation function for the pumping mode.} An example
of the behavior of the second-order correlation function $G^{(2)}_1(\tau)$
in the weak-pumping limit is shown in Fig.~\ref{fig32}. There, the
parameter values $\Gamma=0.07\kappa_0$, $\kappa_1=20\kappa_0$, and
$N=3.9$ were used. The order of magnitude of $G^{(2)}_1(0)$ as well as
the temporal decay rate of $G^{(2)}_1$ (which is much larger than
the decay rate of $G^{(2)}_0$) are in good agreement with the analytical
predictions. It should be noted that the time evolution of the mean
population of the pumping mode $\langle a^{\dagger}_1 a_1(t) \rangle$
starting from an initial vacuum state contains both characteristic time
scales $1/q_0$ and $1/q_1$. This is quite different from the behavior of
$G^{(2)}_1$ which is characterized by $1/q_1$ alone. In the strong
pumping limit the time dependence of $G^{(2)}_1$ could again not be
recognized due to numerical noise. This is consistent with the fact that
the linearized fluctuation analysis predicts a small value of
$|g^{(2)}_1(0)-1|$ for the parameter values investigated. Furthermore, the
calculations
indicated that Eq.~(\ref{ssn1}) ceases to be valid for large $\Gamma$.
It was not possible to reduce $\bar{n}_1$ to very small values
in which case a more pronounced anti-bunching would be expected.

In conclusion, we see that the results obtained from the linear
fluctuation analysis describe well the essential aspects of the behavior
of the correlation functions $D^{(1)}_0$ and $G^{(2)}_0$ and may be used
as a first quantitative estimate.

\section{Conclusion and outlook}

In this paper, we have adopted an operational approach to introduce several
classes of coherence of the Schr\"odinger field. Of particular importance
are density coherence, which is connected to far-off resonance imaging
measurements, and field coherence, which can be measured in
ionization-type measurements. One can readily imagine further classes of
coherence associated with other types of measurements, but they are probably
not as important as field and density coherence.

One question of considerable significance in the future will be to quantify
the usefulness of various sources of Schr\"odinger fields for specific
applications. In optics, one of the most important characteristic of lasers
is their spectral width, and higher-order coherence plays a limited role
in most cases. By analogy, past studies of atom lasers have
concentrated on their spectral width, as determined by their first-order
field correlations. It as been found that this linewidth can be quite
broad, specially in the presence of elastic collisions. Indeed, things can
be so bad that the atom laser linewidth is broader than the natural linewidth
of the atom cavity, in sharp contrast to the optical situation where, of
course, the reverse is true. It is not clear however whether this is a
useful way to determine the quality and usefulness of an atom laser. While
this is likely to be the case in some interferometric applications, other
possible uses of atom laser beams, such as coherent lithography, may well
require only a high degree of density coherence, in which case, as we have
seen, elastic collisions do not play a detrimental role. Hence, we believe 
that it is important at this point to start analyzing in detail the 
coherence requirements of specific atom laser applications, so as to 
optimize their design.

\acknowledgements
This work is supported in part by the U.S. Office of Naval Research
Contract No. 14-91-J1205, by the National Science Foundation Grant
PHY95-07639, by the U.S. Army Research Office and by the
Joint Services Optics Program.

\end{document}